\documentclass[aps,pre,twocolumn,superscriptaddress,showpacs]{revtex4}

\usepackage{amssymb}
\usepackage{epsfig}
\usepackage{amsmath}
\usepackage{times}
\begin{document}

\title{Optimization of synchronization in gradient clustered networks}

\author{Xingang Wang}
\affiliation{Temasek Laboratories, National University of Singapore, Singapore, 117508}
\affiliation{Beijing-Hong Kong-Singapore Joint Centre for Nonlinear \& Complex Systems
(Singapore), National University of Singapore, Kent Ridge, Singapore, 119260}
\author{Liang Huang}
\affiliation{Department of Electrical Engineering, Arizona State University, Tempe,
Arizona 85287, USA}
\author{Ying-Cheng Lai}
\affiliation{Department of Electrical Engineering, Arizona State University, Tempe,
Arizona 85287, USA}
\author{Choy Heng Lai}
\affiliation{Beijing-Hong Kong-Singapore Joint Centre for Nonlinear \& Complex Systems
(Singapore), National University of Singapore, Kent Ridge, Singapore, 119260}
\affiliation{Department of Physics, National University of Singapore, Singapore, 117542}

\begin{abstract}

We consider complex clustered networks with a gradient structure,
where sizes of the clusters are distributed unevenly. Such
networks describe more closely actual networks in biophysical
systems and in technological applications than previous models.
Theoretical analysis predicts that the network synchronizability
can be optimized by the strength of the gradient field but only
when the gradient field points from large to small clusters. A
remarkable finding is that, if the gradient field is sufficiently
strong, synchronizability of the network is mainly determined by
the properties of the subnetworks in the two largest clusters.
These results are verified by numerical eigenvalue analysis and by
direct simulation of synchronization dynamics on
coupled-oscillator networks.

\end{abstract}

\date{\today }
\pacs{05.45.Xt, 05.45.Ra, 89.75.Hc}
\maketitle

It has been recognized in biological physics that at the cellular
level, information vital to the functioning of the cell is often
processed on various networks with complex topologies
\cite{Cell_CM_Overview}. At a systems level, organizing
information using the network idea has also become fundamental to
understanding various biological functions. A key organizational
feature in many biological systems is the {\em clustered}
structure where biophysical and biochemical interactions occur at
a hierarchy of levels. Examples include various protein-protein
interaction networks \cite{Ho:2002,SM:2003} and metabolic networks
\cite{RSMOB:2002}. In biology and network science, a fundamental
issue is synchronization
\cite{Strogatz:2003,Network_Synchronization}. The aim of this
paper is to study synchronization in clustered complex networks
with uneven cluster-size distribution and asymmetrical coupling.
Since this type of network structure is also important to physical
and technological systems such as electronic-circuit networks and
computer networks \cite{Milo:2002,VPV:2002,ESMS:2003},
understanding synchronization in such networks will be of broad
interest.

There has been recent effort to study synchronization in complex
clustered networks \cite{PLGK:2006,ADP:2006}. A general assumption
in these works is that all clusters in a network are on the equal
footing in the sense that their sizes are identical and the
interactions between any pair of clusters are symmetrical. In
realistic applications the distribution of the cluster size can be
highly uneven. For example, in a clustered network with a
hierarchical structure, the size of a cluster can in general
depend on the particular hierarchy to which it belong. More
importantly, the interactions between clusters in different
hierarchies can be highly asymmetrical. For instance, the coupling
from a cluster at a top hierarchy to a cluster in a lower
hierarchy can be much stronger than the other way around. An
asymmetrically interacting network can in general be regarded as
the superposition of a symmetrically coupled network and a
directed network, both being weighted. A weighted, directed
network is a {\em gradient network} \cite{TB:2004,PLZY:2005}, a
class of networks for which the interactions or couplings among
nodes are governed by a gradient field. Our interest is then the
synchronizability and the actual synchronous dynamics on complex
clustered networks with a gradient structure.

For a complex gradient clustered network, a key parameter is the
strength of the gradient field between the clusters, denoted by
$g$. A central issue is how the network synchronizability depends
on $g$. As $g$ is increased, the interactions among various
clusters in the network become more directed. From a
dynamical-system point of view, uni-directionally coupled systems
often possess strong synchronizability \cite{NM:2006,WLL:2007}.
Thus, intuitively, we expect to observe enhancement of the network
synchronizability with the increase of $g$. The question is
whether there exists an optimal value of $g$ for which the network
synchronizability can be maximized. This is in fact the problem of
optimizing synchronization in clustered gradient networks, and our
findings suggest an affirmative answer to the question. In
particular, we are able to obtain solid analytic insights into a
key quantity that determines the network synchronizability. The
theoretical formulas are verified by both numerical eigenvalue
analysis and direct simulation of oscillatory dynamics on the
network. The existence of an optimal state for gradient clustered
networks to achieve synchronization may have broad implications
for evolution of biological networks and for practical
applications such as the design of efficient computer networks.

Our general setting is network with $N$ nodes and $M$ clusters,
where $n_{m}$ is the size of cluster $m$ and $V_{m}$ denotes the
set of nodes it contains ($m=1,...,M$). Each pair of nodes is
connected with probability $p_{s}$ in the same cluster and with
probability $p_{l}$ in different clusters, where $p_{s}>p_{l}$
\cite{PLGK:2006}. For a coupled oscillator network with arbitrary
connecting topology, its synchronizability is determined
\cite{BP:2002} by the interplay between the transverse stability
of the local-node dynamics $\mathbf{F}(\mathbf{x})$ and the
eigenvalue spectrum of the coupling matrix $C$, which can be
sorted conveniently as $\lambda_1=0<\lambda_2 \leqslant \cdots
\leqslant \lambda_N$, where $\lambda_1=0$ underlies the
synchronization solution. A typical nonlinear oscillator in the
synchronization manifold is transversely stable only when some
generalized coupling parameter $\sigma$ falls in a finite range:
$\sigma \in \lbrack \sigma_1,\sigma_2]$, which is determined by
the single-oscillator dynamics. The network is synchronizable if
all the normalized eigenvalues except $\lambda _{1}$ can be
contained within this range: $\sigma _{1}<\varepsilon\lambda
_{2}\leqslant \cdots \leqslant \varepsilon\lambda _{N} < \sigma
_{2}$, where $\varepsilon$ is a specific coupling parameter. For
convenience, we consider the following class of coupled-map
networks:
$\mathbf{x}_{t+1}^{i}=\mathbf{f}(\mathbf{x}_{t}^{i})-\varepsilon
\sum_{j}C_{ij}\mathbf{H}\left[
\mathbf{f}(\mathbf{x}_{t}^{j})\right] $, where
$\mathbf{x}_{t+1}^{i}=\mathbf{f}(\mathbf{x}_{t}^{i})$ is a
$d$-dimensional map representing the local dynamics of node $i$,
$\varepsilon $ is a global coupling parameter, and $\mathbf{H}$ is
a coupling function. The rows of the coupling matrix $C$ have zero
sum to guarantee an exact synchronized solution:
$\mathbf{x}_{t}^{1}=\mathbf{x}_{t}^{2}=...=\mathbf{x}_{t}^{N}=\mathbf{s}_{t}$.
For certain types of oscillator dynamics and coupling functions,
say, for example, the linearly coupled logistic oscillators we are
going to study in the following, $\sigma_{N}$ is sufficiently
large \cite{HYL:1998}. In such cases the condition
$\varepsilon\lambda_N<\sigma_2$ is naturally satisfied and the
synchronizability of network is only determined by $\lambda_{2}$.
For simplicity, we will restrict our study to such types of
oscillator dynamics and coupling functions.

We first develop a theory for networks consisting of two clusters
(the theory can be generalized to multiple-cluster networks).
Without a gradient field, the adjacent matrix $A$ is such that
$A_{ij}=1$ if there is a link between node $i$ and node $j$, and
$A_{ij}=0$ otherwise. To introduce a coupling gradient field from
cluster $1$ to cluster $2$, for each inter-cluster link ($i,j$),
$i\in V_{1}$ and $j\in V_{2}$, we deduce an amount $g$ from
$A_{ij}$ (corresponding to the coupling from node $j$ to node $i$)
and add it to $A_{ji}$ so that the total coupling strength is
conserved. In this sense the gradient field can be said to point
from cluster 1 to cluster 2. The coupling matrix $C$ is defined as
$C_{ij}=-A_{ij}/k_i$, where $k_i=\sum_{j=1}^{N}A_{ij}$ is the
weighted degree of node $i$, and $C_{ii}=1$.

The eigenvalue spectra of $C$ and of its transpose $C^T$ are
identical. Let
$\hat{e}_{2}=(e_{1},e_{2},...,e_{n_{1}},e_{n_{1}+1},...,e_{N})^{T}$
be the normalized eigenvector associated with $\lambda _{2}$ of
$C^T$. Since $\sum_{j=1}^{N}C^T_{j,i}=\sum_{j=1}^{N}C_{ij}=0$, the
eigenvectors associated with non-zero eigenvalues of $C^T$ have
zero sum: $\sum_{j=1}^{N}\hat{e}_{2,j}=0$ \cite{zerosum}. From
$C^T\hat{e}_{2}=\lambda _{2}\hat{e}_{2}$ we have $\lambda
_{2}=\hat{e}_{2}^{T}C^T\hat{e}_{2}=\sum_{i,j=1}^{N}e_{i}C_{ij}e_{j}$.
For a clustered network, the elements in $\hat{e}_{2}$ have a
special distribution: $e_{i}\approx E_{1}$ for $i\in V_{1}$ and
$e_{j}\approx E_{2}$ for $j\in V_{2}$ \cite{PLGK:2006}, where the
two constant values $E_{1}$ and $E_{2}$ can be obtained from the
normalization condition $\hat{e}_{2}^{T}\hat{e}_{2}=1$ and the
zero-sum property. We obtain
$E_{1}=-\sqrt{n_{2}/(n_{1}n_{2}+n_{1}^{2})}$ and
$E_{2}=\sqrt{n_{1}/(n_{1}n_{2}+n_{2}^{2})}$ (the signs of $E_{1}$
and $E_{2}$ are interchangeable since $E_{1}E_{2}<0$). This can
greatly simplify the calculation of $\lambda _{2}$, which now can
be written as $\lambda _{2}\approx
\sum_{i=1}^{N}e_{i}\{(C_{i1}+C_{i2}+...+C_{in_{1}})E_{1}+
(C_{in_{1}+1}+C_{in_{1}+2}+...+C_{iN})E_{2}\}$. The non-zero
elements in $C$ can be calculated as follows. For $i \in V_{1}$,
$k_{i}\approx n_{1}p_{s}+n_{2}p_{l}(1-g)$, if $j \in V_{1}$,
$C_{ij}=-1/k_i\equiv g_{11}$, and there are approximately
$n_{1}p_{s}$ non-zero elements for each $i$. If $j \in V_{2}$, we
have $C_{ij}=-(1-g)/k_i\equiv g_{12}$. For $i \in V_{2}$,
$k_{i}\approx n_{2}p_{s}+n_{1}p_{l}(1+g)$, if $j \in V_{1}$,
$C_{ij}=-(1+g)/k_i\equiv g_{21}$ and, if $j \in V_{2}$,
$C_{ij}=-1/k_i\equiv g_{22}$. Since $C_{ii}=1$, the calculation
can be further simplified as $\lambda _{2}\approx
\sum_{i=1}^{n_{1}}e_{i}\{E_{1}+g_{11}E_{1}n_{1}p_{s}+
g_{12}E_{2}n_{2}p_{l}\}+\sum_{i=n_{1}+1}^{N}e_{i}\{g_{21}E_{1}n_{1}p_{l}+
E_{2}+g_{22}E_{2}n_{2}p_{s}\}$. Using
$\sum_{i=1}^{n_{1}}e_{i}\approx n_1E_1$,
$\sum_{i=n_{1}+1}^{N}e_{i}\approx n_2E_2$ and
$n_1E_1^2+n_2E_2^2=1$ (the normalization condition), we obtain
\begin{equation} \label{eq:2_cluster_lambda2}
\lambda_{2}=1+(E_{1}^{2}n_{1}^{2}g_{11}+E_{2}^{2}n_{2}^{2}g_{22})p_{s}
+E_{1}E_{2}n_{1}n_{2}p_{l}(g_{12}+g_{21}).
\end{equation}
In Eq. (\ref{eq:2_cluster_lambda2}), the unity comes from the
diagonal elements in $C$, it defines the upper limit for
$\lambda_2$ (this special case is associated with one-way coupled
tree-structure networks \cite{NM:2006,WLL:2007}). The second term
is contributed by the intra-connection of cluster $1$ and cluster
$2$. The last term corresponds to the inter-connection between the
clusters. The parameter $g$ is contained in these terms via
$g_{ij}$. For a given 2-cluster network, the optimal gradient
strength $g_0$ that maximizes $\lambda_2$ can be determined by
setting $\partial \lambda _{2}/\partial g=0$, which gives
\begin{equation} \label{eq:optimal}
g_{o}=\frac{2n_1-N}{Np_l}(p_s-p_l).
\end{equation}
(Please note that in deriving $g_0$ we actually get two such
values: $g_0$ and $g_{0}^{'}=N(p_s+p_l)/[(N-2n_1)p_l]<-1$. Since
in our network model $|g|$ is defined within range $[0,1]$, the
value $g_{o}^{'}$ is therefore discarded.)

\begin{figure}[tbp]
\begin{center}
\epsfig{figure=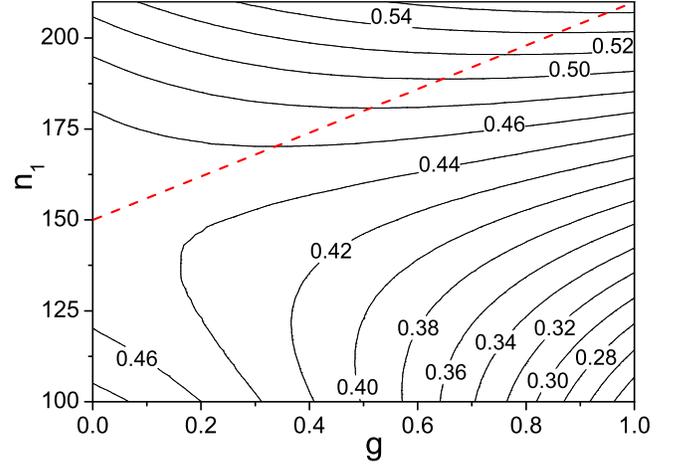,width=\linewidth} \caption{(Color
online) Theoretical contour plot of $\lambda_2$ in the $(g,n_1)$
plane, for a $2$-cluster network of $n_1 + n_2 = 300$ nodes. Other
parameters are $p_{l}=0.2$ and $p_{s}=0.7$. The dashed line is
given by Eq. (\ref{eq:optimal}), which determines, for fixed value
of $n_1$, the optimal gradient strength $g_0$.}
\label{fig:coutour} \label{fig:lambda_2_contour}
\end{center}
\end{figure}

Equation (\ref{eq:2_cluster_lambda2}) reveals some interesting
features about the dependence of $\lambda_2$ on key parameters of
the clustered network. To give an example, we show in Fig.
\ref{fig:lambda_2_contour} a contour plot of $\lambda_2$,
calculated using the theoretical formula Eq.
(\ref{eq:2_cluster_lambda2}), in the parameter plane spanned by
$n_1$ and $g$, where $n_1 + n_2 = 300$. It gives, for fixed value
of $n_1$, the dependence of $\lambda_2$ on gradient strength.
Since, by our construction, the gradient field points from cluster
1 to cluster 2, the upper half region ($n_1 > 150$) in Fig.
\ref{fig:lambda_2_contour} represents gradient clustered networks
for which the gradient field points from the large to the small
cluster. For any network defined in this region, for any fixed
value of $g$, $\lambda_2$ increases monotonically with $n_1$,
indicating enhanced network synchronizability with the size of the
large cluster. However, for a fixed value of $n_1$, $\lambda_2$
first increases, reaches maximum for some optimal value of $g
\equiv g_0$, and then decreases with $g$. The dependence of $g_0$
on $n_1$ is revealed by the dashed line in the figure [Eq.
(\ref{eq:optimal})]. We see that, when the gradient field is set
to point from the large to the smaller cluster, in order to
optimize the network synchronizability, larger gradient strength
is needed for larger difference in the cluster sizes. In contrast,
in the lower-half of Fig. \ref{fig:lambda_2_contour} where $n_1 <
n_2$, $\lambda_2$ tend to decrease as $g$ is increased (for fixed
$n_1$) or when the difference between the sizes of the two
clusters enlarges. This indicates that, when the gradient points
from the smaller to the larger cluster, the network
synchronizability continuously weakens as the the gradient field
is strengthened.

\begin{figure}[tbp]
\begin{center}
\epsfig{figure=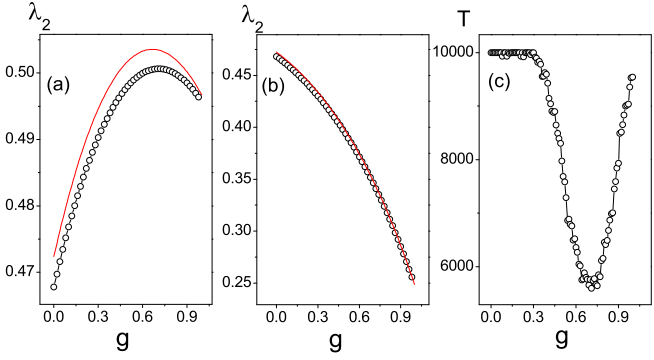,width=\linewidth} \caption{(Color
online) For a gradient network of two clusters with $N = 300$
nodes, numerically obtained (circles) dependence of
$\protect\lambda _{2}$ on the strength $g$ of the gradient field
for the two cases where (a) the gradient field points from the
larger to the small cluster ($n_1 = 190 > N/2$) and (b) the
opposite ($n_1 = 110 < N/2$). The solid curves are from theory.
(c) For $n_1 = 190$, actual synchronization time versus $g$ for a
clustered network of chaotic logistic maps. We observe a sharp
reduction in the time as $g$ approaches its optimal value,
indicating a stronger synchronizability. Other parameters are
$p_l=0.2$, $p_s=0.7$. Each point is the average of 100 random
realizations.} \label{fig:simulation}
\end{center}
\end{figure}

To provide support for our theoretical formula Eq.
(\ref{eq:2_cluster_lambda2}), we consider the same network in Fig.
\ref{fig:simulation} and directly calculate the eigenvalue
spectrum for a systematically varying set of values of $g$. Figure
\ref{fig:simulation}(a) shows $\lambda_2$ versus $g$ (open
circles) for the case where the gradient field points from the
large to the small cluster ($n_1 = 190 > N/2$) and Fig.
\ref{fig:simulation}(b) is for an opposite case ($n_1 = 110 <
N/2$). The solid curves are theoretical predictions. We observe a
good agreement. To gain insight into the actual dynamics of
synchronization on the network, we use the logistic map
$f(x)=4x(1-x)$ as the local dynamics, $\varepsilon = 1$, and
choose $\mathbf{H}(\mathbf{x})=x$ as the coupling function. For
the logistic map, we have $\sigma _{1}=0.5$, $\sigma _{2}=1.5$
\cite{JJ:2002}. We find numerically $\lambda_N \approx 1.1 <
\sigma_2$. Thus the synchronization condition becomes $\lambda
_{2}>\sigma _{1}=0.5$. We have calculated the average
synchronization time $T$ as a function of $g$, where $T$ is the
time needed to reach $\sum_{i=1}^{N}\left\vert (x^{i}-\langle x
\rangle)\right\vert /N<\delta =10^{-5}$ and $\langle x \rangle
\equiv \sum_{i=1}^{N}x^{i}/N$ (the system is considered as
unsynchronizable when $T>10^{4}$). As $g$ approaches the optimal
value $g_0$, we observe a sharp decrease in $T$, as shown in Fig.
\ref{fig:simulation}(c), indicating a significant enhancement of
the network synchronizability. After reaching the minimum at
$g_0$, the time increases as $g$ is increased further, as
predicted by theory.

The theory we have developed for two-cluster networks can be
extended to multiple-cluster networks. Consider a $M$-cluster
network, where each cluster contains a random subnetwork. Assume
the size of the clusters satisfy $n_1>n_2>n_3\geqslant \cdots
\geqslant n_M$, a coupling gradient field can be defined as for
the two-cluster case. For a random clustered network, the weighted
degree can be written as $k_i\approx \sum_{j=1}^{N}A_{ij}=n_m p_s
+ (N-n_m) p_l + p_{l}g (\sum_{l,n_m<n_l}n_{l}-\sum_{l^{\prime
},n_m>n_{l^{\prime }}}n_{l^{\prime }})\equiv K_m$. Define $g_{ml}$
as the average value of the non-diagonal, non-zero elements
$C_{ij}$. For $i \in V_m$ and $j \in V_l$, we have
$g_{mm}=-1/K_m$, $g_{ml}=-(1-g)/K_m$ for $n_m>n_l$,
$g_{ml}=-(1+g)/K_m$ for $n_m<n_l$, and $g_{ml}=-1/K_m$ for
$n_m=n_l$. For the second eigenvector of $C^T$, e.g. $C^T
\hat{e}_{2}=\lambda_2 \hat{e}_{2}$, its components have a
clustered structure, i.e., for all $i \in V_m$,
$\hat{e}_{2,i}\approx E_m$ while they may vary significantly for
different clusters. The eigenvalue $\lambda_2$ can then be
expressed as $\lambda_{2} = \hat{e}_{2}^{T}C^T\hat{e}_{2} =
\sum_{i,j=1}^{N}e_{i}C_{ij}e_{j}=\sum_{i=1}^{N}e_{i}\{E_m + E_m
n_m p_s g_{mm} + \sum_{l\neq m}E_{l}n_{l}p_l g_{ml}\} =
\sum_{m=1}^{M}n_m E_{m}^2 + \sum_{m=1}^{M}E_m^2 n_m^2 p_s g_{mm}
+\sum_{l\neq m}E_{m}E_{l}n_{m}n_{l}p_l g_{ml}$. Taking into
consideration the normalization condition
$\hat{e}_{2}^{T}\hat{e}_{2}=1$, we get $\lambda_{2}=1+
\sum_{m=1}^{M}E_m^2 n_m^2 p_s g_{mm} +\sum_{m,l=1; l\neq m }^N
E_{m}E_{l}n_{m}n_{l}p_l g_{ml}$.

For a general multiple-clustered network, it is mathematically
difficult to obtain an analytic formula for the quantity $E_m$.
However, $E_m$ can be determined numerically. Once this is done,
the general dependence of $\lambda_2$ on $g$ and subsequentially
the optimal gradient strength $g_0$ can be obtained.
In some particular cases, explicit formulas for $E_m$ and
$\lambda_2$ can be obtained. Focusing on the role of the gradient
in determining the synchronizability, we consider the extreme
gradient case: $g=1$. Numerically, we find that for this case,
with respect to the second eigenvector $\hat{e}_{2}$, only $E_1$
and $E_2$ (corresponding to the largest and the second largest
clusters) have non-zero values, while for all $m>2$, $E_m=0$. From
the normalization condition $\hat{e}_{2}^{T}\hat{e}_{2}=1$ and the
zero-sum property $\sum_{j=1}^{N}\hat{e}_{2,j}=0$ (since
$\sum_{j=1}^{N}C_{ij}=0$), we can solve for $E_1$ and $E_2$ as
$E_{1}=-\sqrt{n_{2}/(n_{1}n_{2}+n_{1}^{2})}$ and
$E_{2}=\sqrt{n_{1}/(n_{1}n_{2}+n_{2}^{2})}$. Noticing $g_{12}=0$,
we finally obtain
\begin{eqnarray} \label{eq:M_cluster_lambda_2_2}
\lambda_{2}&=&1+ \sum_{m=1}^{2}E_m^2 n_m^2 p_s g_{mm}
+\sum_{m,l,l\neq m}^2 E_{m}E_{l}n_{m}n_{l}p_l g_{ml}\notag \\
&=&1+ (E_1^2 n_1^2 g_{11}+E_2^2 n_2^2 g_{22}) p_s
+E_{1}E_{2}n_{1}n_{2}p_l g_{21}.
\end{eqnarray}
A numerical verification of Eq. (\ref{eq:M_cluster_lambda_2_2}) is
provided in Fig. 3(a). An observation is that, except for the
difference in $g_{ij}$, Eq. (\ref{eq:M_cluster_lambda_2_2}) has
the same form as Eq. (\ref{eq:2_cluster_lambda2}), indicating that
$\lambda_2$ is mainly determined by the first two largest clusters
and it has little dependence on the details of size distributions
of the remaining clusters. The remarkable implication is that, for
different gradient clustered networks, regardless of the detailed
form of the cluster size distribution, insofar as the two dominant
clusters have similar properties, all networks possess nearly
identical synchronizability.

\begin{figure}[tbp]
\begin{center}
\epsfig{figure=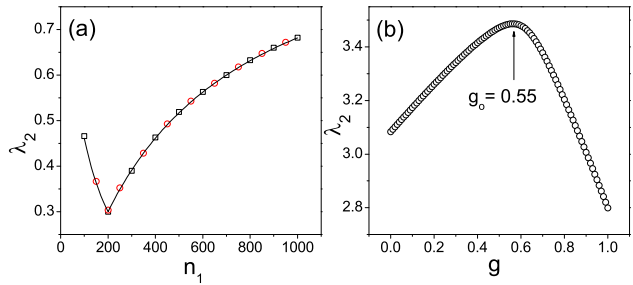,width=\linewidth} \caption{(Color
online) (a) For a 5-cluster network (circles) and a 10-cluster
network (squares), $\lambda_2$ versus $n_1$, the size of the
largest cluster. The solid curve is from theory [Eq.
(\ref{eq:M_cluster_lambda_2_2})]. For the 5-cluster network, the
size of the remaining clusters are $n_2=200$, $n_3=50$, $n_4=30$,
$n_5=20$. For the 10-cluster network, we have $n_2=200$, $n_3$ to
$n_{10}$ are $90$, $80$, $70$, $60$, $50$, $40$, $30$, $20$,
respectively. Other parameters are $p_{l}=0.15$ and $p_{s}=0.7$.
For $n_1 < n_2$, the gradient is actually from cluster 2 to
cluster 1. Each point is the average result of $100$ network
realizations. (b) For a ``cortico-cortical network" of the cat
brain, numerical results of the dependence of $\lambda_2$ on
gradient strength $g$. Synchronization is optimized for $g_0
\approx 0.55$} \label{fig:M_cluster}
\end{center}
\end{figure}

The model of gradient clustered network we have investigated here
is different to the asymmetrical network models in literature. In
Ref. \cite{Network_Synchronization,NM:2006,WLL:2007}, asymmetrical
couplings have been employed to improve network synchronization
and it is found that, for \emph{non-clustered networks},
synchronization is optimized when all nodes are one-way coupled
and the network has a tree-structure \cite{NM:2006}. Different to
this, in our model asymmetrical couplings are only introduced to
inter-cluster links, while couplings on intra-cluster links are
still symmetrical. This special coupling scheme induces some new
properties to the functions of the gradient. Firstly, increase of
gradient will not monotonically enhance synchronization. That is,
directed coupling between clusters, i.e. $g=1$, is not always the
best choice for synchronization. In many cases the optimal
gradient stregth $g_0$ is some value between $0$ and $1$, while
the exact value is determined by the other network parameters
[Eqs. (\ref{eq:2_cluster_lambda2},\ref{eq:M_cluster_lambda_2_2})].
Secondly, the direction of gradient can not be arranged randomly,
it should be always pointing from large to small clusters.
Finally, in the case of $g=1$, network synchronizability is still
related to the network topology, i.e. by the topology of the first
two largest clusters; while for non-clustered network,
synchronizability is only determined by the local dynamics
\cite{NM:2006}.

Can synchronization optimization be expected in realistic
networks? To address this question, we have tested the
synchronizability of a ``cortico-cortical network" of cat brain,
which comprises 53 cortex areas and about 830 fiber connections of
different axon densities \cite{SCANNELL:1999}. The random and
small-world properties of this network, as well as its
hierarchical structure, have been established in several previous
papers \cite{CATNET:PROPERTIES}. According to their functions, the
cortex areas are grouped into 4 divisions of variant size: 16
areas in the visual division, 7 areas in the auditory division, 16
areas in the somato-motor division, and 14 areas in the
frontolimbic division. Also, by the order of size, these divisions
are hierarchically organized \cite{SCANNELL:1999}. With the same
gradient strategy as for the theoretical model, we plot in Fig.
3(b) the variation of $\lambda_2$ as a function of the gradient
strength. Synchronization is optimized at gradient strength about
$g_{o} \approx 0.55$. An interesting finding is that the actual
average gradient of the real network, $g_{ave} \approx 0.37$
\cite{Optimization:real}, is deviating from the optimal gradient
$g_{o}$, indicating a strong but non-optimized synchronization in
healthy cat brain.

While our theory predicts the existence of a gradient field for
optimizing the synchronizability of a complex clustered network,
we emphasize that the actual value of the optimal gradient field
may or may not be achieved for realistic networked systems. Due to
the sophisticated procedure involved to determine the optimal
gradient strength and the actual value for a given network, their
numerical values can contain substantial uncertainties. 
A reasonable test should involve a large scale comparison across
many networks of relatively similar type (say, many different
animals), hopefully demonstrating some kind of correlation between
the optimum gradient and the observed values. Furthermore, such a
test would include a sense of how large the difference between the
optimum and observed is. Due to the current unavailability of any
reasonable number of realistic complex, gradient, and clustered
networks, it is not feasible to conduct a systematic test of our
theory. (As a matter of fact, we are able to find only one
real-world example of gradient clustered network, the cat-brain
network that we have utilized here.) It is our hope that, as
network science develops and more realistic network examples are
available, our theory and its actual relevance can be tested on a
more solid ground.

In summary, we have uncovered a phenomenon in the synchronization
of gradient clustered networks with uneven distribution of cluster
sizes: the network synchronizability can be enhanced by
strengthening the gradient field, but the enhancement can be
achieved only when the gradient field points from large to small
clusters. We have obtained a full analytic theory for gradient
networks with two clusters, and have extended the theory to
networks with arbitrary number of clusters in some special but
meaningful cases. For a multiple-cluster network, a remarkable
phenomenon is that, if the gradient field is sufficiently strong,
the network synchronizability is determined by the largest two
clusters, regardless of details such as the actual number of
clusters in the network. These results can provide insights into
biological systems in terms of their organization and dynamics,
where complex clustered networks arise at both the cellular and
systems levels. Our findings can also be useful for optimizing the
performance of technological networks such as large-scale computer
networks for parallel processing.

XGW acknowledges the great hospitality of Arizona State
University, where part of the work was done during a visit. YCL
and LH were supported by NSF under Grant No. ITR-0312131 and by
AFOSR under Grant No. FA9550-06-1-0024, and No. FA9550-07-1-0045.


\begin{references}
\bibitem{Optimization:real}
Denoting $g_{ij}$ and $g_{ji}$ the weights of the directed
couplings on inter-cluster link $(i,j)$, $i \in V_i$ and $j \in
V_j$, and $n_i$ ($n_j$) the size for cluster $V_i$ ($V_j$). The
average gradient of the cat brain network is calculated as
$g_{ave}=\frac{1}{L}\sum_{l}^{L}(g_{ij}-g_{ji})$, with $L$ the
number of inter-cluster links.
\end{references}
\end{document}